**Dirac fermion reflector by ballistic graphene sawtooth-shaped npn junctions**

Sei Morikawa[1], Quentin Wilmart[2], Satoru Masubuchi[1], Kenji Watanabe[3], Takashi Taniguchi[3], Bernard Plaçais[2], and Tomoki Machida[1,4,a]

[1]*Institute of Industrial Science, University of Tokyo, 4-6-1 Komaba, Meguro, Tokyo 153-8505, Japan*
[2]*Laboratoire Pierre Aigrain, ENS-CNRS UMR 8551, Universités P. et M. Curie and Paris-Diderot, 24, rue Lhomond, 75231 Paris Cedex 05, France*
[3]*National Institute for Materials Science, 1-1 Namiki, Tsukuba 305-0044, Japan*
[4]*Institute for Nano Quantum Information Electronics, University of Tokyo, 4-6-1 Komaba, Meguro, Tokyo 153-8505, Japan*

**Abstract:**

We have realized a Dirac fermion reflector in graphene by controlling the ballistic carrier trajectory in a sawtooth-shaped npn junction. When the carrier density in the inner p-region is much larger than that in the outer n-regions, the first straight np interface works as a collimator and the collimated ballistic carriers can be totally reflected at the second zigzag pn interface. We observed clear resistance enhancement around the np$^+$n regime, which is in good agreement with the numerical simulation. The tunable reflectance of ballistic carriers could be an elementary and important step for realizing ultrahigh-mobility graphene field effect transistors utilizing Dirac fermion optics in the near future.

---

[a] E-mail: tmachida@iis.u-tokyo.ac.jp

# 1. Introduction

The development of fabrication methods for high-mobility graphene/h-BN[1–5] has recently unveiled ballistic carrier-transport phenomena in graphene such as negative bend resistance[4,6], the magnetic focusing effect[7–9], and a magnetoresistance peak due to boundary scattering[10]. Under such a ballistic carrier transport regime, the transmission of carriers can be described by analogy to the geometrical optics of light, opening up a research field of "Dirac fermion optics"[11]. Indeed, by using reflection and refraction of ballistic Dirac fermions at an interface of different carrier-density regions, especially at a pn interface, numerous characteristic functionalities have been realized experimentally, such as waveguides[12], Veselago lenses[13], beam splitters[14], and edge-channel interferometers[15]. More recently, systematic experimental studies on Snell's law at ballistic graphene pn junctions were reported[16]. In these systems, the pn interface can be utilized as an optical component between different refractive-index regions, e.g., lens, mirror, or prism.

Ballistic carrier transport has also been proposed as a tool to turn currents ON and OFF in graphene without inducing an energy gap by using tilted pn junctions[17–22]. First, some groups proposed the graphene electro-optic transistor by a single tilted junction[17–19], and then Sajjad *et al*. proposed more efficient way by combining the collimation and total reflection of ballistic carriers using two sequential pn junctions (npn junctions) with different tilted angle for suppressing ballistic carrier transmission in graphene[20]. After that, Wilmart *et al*. analyzed more detailed performance of this tilted npn junctions with high-bias characteristics by the non-equilibrium Green's function (NEGF) methods[22]. The method has advantages for microwave electronics because the OFF state can be realized at a high-carrier-density region without degrading graphene's carrier mobility by using Dirac fermion optics, which is totally different from conventional methods such as the creation of nanoribbons[23], addition of embedding defects[24], and use of bilayer graphene[25,26].

In this paper, we experimentally demonstrate resistance enhancement in sawtooth-shaped npn junctions by using the collimation and total reflection of ballistic carriers. The tunable reflectance of ballistic carriers is a key ingredient of field effect transistors (FETs) that had been proposed[17–20,22], and it could be an elementary and important step for realizing ultrahigh-mobility graphene FETs utilizing Dirac fermion optics.

## 2. Theory and methodology

First, we consider an interface between two regions of graphene with different carrier densities $n_1$ and $n_2$ [Fig. 1(a)]. The flow of ballistic carriers in the two regions can be regarded as the propagation of electron waves whose wave vectors are $(|n_1|\pi)^{0.5}$ and $(|n_2|\pi)^{0.5}$, respectively. Analogously to light in the geometrical optics regime, the incident ballistic carrier wave is refracted according to Snell's law[27],

$$\sin\theta_2 = \text{sgn}(n_1 n_2)(|n_1/n_2|)^{0.5}\sin\theta_1.$$

The refraction angle $|\theta_2|$ is smaller (larger) than the incident angle $|\theta_1|$ when $|n_1| < |n_2|$ ($|n_1| > |n_2|$). Even negative refraction can occur if $n_1 n_2 < 0$[27].

Next, we consider a graphene in-plane junction whose carrier densities are $n_1$, $n_2$, and $n_1$, respectively, as shown in Figs. 1(b) and 1(c); the right interface is tilted from the left interface by an angle of $\alpha$. When $|n_2| > |n_1|$ is satisfied, the flow of ballistic carriers injected to the first straight interface [left interface in Figs. 1(b) and 1(c)] from an unrestricted incident angle is collimated to a limited refraction angle between $-\theta_c$ and $\theta_c$ after the transmission through the interface, where the critical angle is given by $\theta_c =$ Arcsin$(|n_1|/|n_2|)^{0.5}$. Thus, the first interface can be regarded as a collimator for the electron wave in graphene. Then, the electron wave collimated within the spread angle $2\theta_c$ is transmitted to the second interface [right interface in Figs. 1(b) and 1(c)] with an incident angle between $-\theta_c + \alpha$ and $\theta_c + \alpha$. If all the incident angles are larger than $\theta_c$, i.e., $2\theta_c < \alpha$, all the incident ballistic carriers are totally reflected at the second interface. This condition can be realized when $|n_2|$ is much larger than $|n_1|$ because $\theta_c =$ Arcsin$(|n_1|/|n_2|)^{0.5}$. In this case, the transmission of ballistic carriers is completely suppressed; thus, the situation can be regarded as an OFF state in the FET operation. In addition, since the critical angle $\theta_c$ can be controlled experimentally by changing the ratio of carrier densities in the two regions according to $\theta_c =$ Arcsin$(|n_1|/|n_2|)^{0.5}$, we can switch between the OFF state [Fig. 1(b)] and the ON state [Fig. 1(c)] electrostatically. The advantage of this concept for microwave electronics is that we can realize the OFF state even by *increasing* the carrier density $|n_2|$. This implies that we can use this device while keeping the regime with large carrier density, which benefit the dynamical and noise properties of the device due to the intrinsic transmission[17]. Such concept is totally different from the normal method realizing the OFF state by decreasing the carrier density, around Dirac point in case of graphene.

We also extend our model to include the effect of the split junction in a realistic device, considering an angle distribution of transmission probability across the junction $T(\phi)$[17,28–30], which can be derived from equation (10) in Cayssol *et al.*[31]. $\phi$ is the incident angle of ballistic carriers to the junction. In particular, when the potential profile across the junction is smooth owing to the finite thickness of the top gate dielectric, $T(\phi)$ is much sharper than in an abrupt junction with strong angular filtering similar to Gaussian one, as shown in Fig. 1(d). Moreover, $T(\phi)$ in a pn junction is significantly suppressed except at $\phi \sim 0$, even if $\phi < \theta_c$ [Fig. 1(e)], compared to that in an nn junction, as shown in Fig. 1(e). Such results indicate that, in smooth npn junctions with a sawtooth-shaped gate, the collimation effect at first interface and filtering effect at second interface become more prominent, and the OFF current can be suppressed more than that in sharp and/or nnn junctions, although the ON current is also suppressed[20,22].

By using such $T(\phi)$, we used the scattering theory as in Wilmart *et al.*[22], rather than NEGF method, because NEGF requires too much computational resources; we numerically calculated the resistance across one simple triangular-shaped ballistic region with carrier density $n_2$ surrounded by a region with carrier density $n_1$, as shown in the inset of Fig. 1(f). The larger the number of reflections we considered, the more the beam intensity inside the junctions decreases[22]. Therefore, in order to truncate the residual electrons inside the junction reasonably, we considered ten internal reflections of ballistic carriers in the triangular region, which is a sufficient number since we have checked that this truncation does not affect reflection data by more than 10%. Following the experimental parameters from our devices examined below, we set the tilt angle $\alpha$ to 20° and the length of the junction $d$ to correspond to a parameter of the steepness of the carrier density gradient, 120 nm, which was derived numerically using the finite element method. Figure 1(f) shows the result of resistance as a function of $n_1$ and $n_2$. In addition to the charge neutrality points, $n_1$

= 0 and $n_2$ = 0, we can see the resistance enhancement at the np$^+$n, pn$^+$p, nn$^+$n, and pp$^+$p regions ($|n_2| \gg |n_1|$), which is consistent with our prediction that the reflection of ballistic carriers can be enhanced when $|n_2|$ is much larger than $|n_1|$. Note that the resistance enhancement around the bipolar regime is larger than that for the unipolar regime, which is consistent with the transmission probability difference shown in Fig. 1(e).

## 3. Device fabrication and characterization

In order to realize this ballistic device experimentally, we fabricated high-mobility dual-gated graphene by encapsulating graphene with h-BN crystals[32,33] using the "pick-up" transfer method[1]. First, we assembled h-BN/graphene/h-BN van der Waals heterostructures and attached Cr/Pd/Au one-dimensional contacts[1], following the same procedure used in our previous report, from which we can estimate that the expected carrier mobility is greater than 50,000 cm$^2$/Vs and the mean free path is on the order of 1 μm[8]. Then, we fabricated a metal top gate by using the top 50-nm-thick h-BN as a dielectric. The metal top gate is sawtooth-shaped, as shown schematically in Fig. 2(a). The sawtooth shape consists of a sequence of triangular junctions, and the elementary unit is a dashed region depicted in Fig. 2(a), embodying our concept in Figs. 1(b) and 1(c), whose tilt angle $\alpha$ is ~20° with a variation of ±5° due to the lithography imperfectness. The influence of this variation on the results will be discussed later. By using the sawtooth shape, i.e., the sequential triangular units, we kept the average local top-gate length $l$ sufficiently short (~ 0.4 μm) because we need to maintain the ballistic regime. An atomic force microscopy (AFM) image of this device is shown in the inset of Fig. 3(a). The channel width $W$ and length $L$ are ($W$, $L$) ~ (2.4, 2.2) μm. We used a conventional lock-in technique and measured the four-terminal resistance $R_{4t}$. By tuning the back-gate and top-gate bias voltages $V_b$ and $V_t$, we can independently control the carrier densities $n_2$ and $n_1$ of the top-gated region and the uncovered region through $n_2 = (C_t V_t + C_b V_b)/e$ and $n_1 = (C_t' V_t + C_b V_b)/e$, as depicted schematically in Fig. 2(b), where $C_b$ is the back-gate capacitance and $C_t$, $C_t'$ are top-gate capacitances for the regions with carrier density $n_2$, $n_1$, respectively. The values of ($C_b$, $C_t$, $C_t'$) = (1.2, 6.0, 0.1) × $10^{-4}$ C/m$^2$ can be derived from the quantum Hall measurements and the slope of the Dirac point trajectories in the ($V_b$, $V_t$) plane[34]. We performed experiments at $T$ = 120 K so that the carrier transport could be maintained in a ballistic regime and coherent resistance oscillations could be suppressed owing to the thermal phase averaging[13], i.e., an experimental condition where the carrier mean free path $L_{mfp}$ > average local top-gate length $l$ > the thermal length $L_T$. We observed a negative resistance in $R_{4t}$ for $n_2 = n_1$ at $T$ = 120 K, as shown in Fig. 3(a), which is qualitatively the same as the earlier observation of the negative bend resistance[6] and can be regarded as a signature of the ballistic carrier transport regime.

## 4. Results and discussion

Figure 2(c) shows experimental results for $R_{4t}$ as a function of $n_2$ and $n_1$. In the vicinity of the np$^+$n and pn$^+$p regime ($|n_2| \gg |n_1|$ and $n_1 n_2 < 0$), $R_{4t}$ clearly shows resistance enhancement, which qualitatively agrees well with our simulation result in Fig. 1(f). We cannot see clear resistance enhancement around the nn$^+$n and pp$^+$p regime ($|n_2| \gg |n_1|$ and $n_1 n_2 > 0$), possibly because the collimation and total reflection effect are so small in this regime, supported by the relatively broad transmission probability distribution in an nn junction as shown in Fig. 1(e), that they could be comparable to the other ballistic effects such as negative bend resistance[6].

Figure 3(a) shows $R_{4t}$ as a function of $n_2$ at fixed $n_1 = 0.9, 1.2, \ldots, 7.8 \times 10^{11}$ cm$^{-2}$ [top to bottom], corresponding to the line cuts of Fig. 2(c). For the npn ($n_2 < 0$) regimes, $R_{4t}$ is strongly enhanced when we increase $|n_2|$. In addition, the dip position of the $R_{4t}$ versus $n_2$ curve, which corresponds to the place where the resistance enhancement starts (as indicated by a purple arrow), moves toward higher $|n_2|$ as we change $|n_1|$ to larger values. These results are qualitatively consistent with our proposal that the ratio of $|n_2|/|n_1|$ is important for resistance manipulation and should be large in order to realize the collimation and total reflection of ballistic carriers. The reproducibility of such resistance enhancement is also checked for another sawtooth-shaped npn junction with $\alpha \sim 20°$, as shown in Fig. 3(c). Even though the maximum carrier density is different from that in Fig. 3(a) owing to the different gate leak voltage, we kept the same length scale of the $x$-axis for the same carrier-density interval for fair comparison between Figs. 3(a) and 3(c). From this comparison, we can see that the resistance enhancement has almost the same slope in $R_{4t}$-$n_2$ plots, which support the reproducibility of the switching effect we discussed. The quantitative performance of this device, characterized by $\Delta R \sim 0.2$ k$\Omega$ and $R_{\text{off}}/R_{\text{on}} \sim 1.3$ from Fig. 3(a), is quite modest for application purposes. We will comment on the additional modifications required to realize more practical performance in the last part of this section.

To compare these experimental results with the simulated results, Fig. 3(b) shows line cuts of Fig. 1(f) for the same value of $n_1$ as in Fig. 3(a). The qualitative behaviors that we discussed above are all well reproduced. In addition, the approximate order of the value of resistance ($\sim 0.6$ k$\Omega$) and the dip position trajectories with increasing $n_1$ (see Appendix A) show good quantitative agreement without any fitting parameter, which indicates the predictive character of our simulation model and the high quality of our device.

To confirm that the resistance enhancement observed in the present work is due to the sawtooth gate junction, the same measurements were carried out in another device with a rectangular junction, as shown in the inset of Fig. 3(d). This device was fabricated from the same h-BN/graphene/h-BN flake and was measured at the same time as the device used in Fig. 3(c). In this device, we did not observe resistance enhancement [Fig. 3(d)] and such qualitative sign difference of resistance slope as a function of $n_t$ between sawtooth and rectangular junctions confirms that the collimation and total reflection of ballistic carriers enhance the resistance in Figs. 3(a) and (c).

Our resistance enhancement, characterized by $\Delta R \sim 0.2$ k$\Omega$ and $R_{off}/R_{on} \sim 1.3$ from Fig. 3(a), is not sufficiently large at this stage. Actually, even between bulk graphene and single pn junction can induce the same order of ON/OFF ratio (>> 1.1 from [20]). However, the combination of collimation and total reflection using npn junctions could potentially induce more suppressed OFF current and even transport gap[20,22], so we suggest three points to improve the performance of this device for the future application. First, we should eliminate the lithography roughness in the sawtooth gate shape and establish the influence of this roughness on the FET performance because geometrical imperfections could lead to spurious reflections or optical aberrations. Second, we should consider the influence of edge scattering and the dependence of FET performance on the tilt angle $\alpha$. From the scattering theory without considering edge scattering[22], $\alpha \sim 45°$ is the most appropriate angle for realizing large resistance enhancement. However, as shown in Appendix B, we cannot observe any larger resistance enhancement in devices with $\alpha \sim 45°$. From other simulation results including the effect of edge scattering, though the gate structure is a little bit different from our device, the *transmission* becomes even maximum at $\alpha \sim 45°$[29]. Therefore, it is a future task for us to find the best value of $\alpha$ from both experimental and theoretical studies in order to realize a much higher resistance enhancement. As another possible factor, we might have to consider the fact that the edge of the sample is not exactly perpendicular to the collimator (first junction), as can be seen in the inset of Fig. 3(a), which could change the influence of edge scattering. Finally, the profile of the local electrostatic potential at the junction should be modified depending on the purpose. As expected from the effect of potential steepness on transmission probability shown in Fig. 1(d), the total transmission should change if we change the steepness of the junction potential. If the device requires a large gain, which implies a large ON current, the potential should be steep. On the other hand, if we need a suppressed OFF current, the potential should be smooth. By changing the thickness of the h-BN dielectric layer, which determines the steepness of electrostatic potential at the junction, we can tune such properties as required.

Since this device has the potential to realize an OFF state at high carrier density region, it could be a breakthrough for the application of graphene FETs in microwave electronics[22] without inducing mobility degradation. In addition, many graphene devices have recently been found to show ballistic effects even at room temperature[6], and we actually observed resistance enhancement up to 280 K (see the temperature dependence and its correlation with carrier mean free path in Appendix C). Such temperature robustness results from the graphene's high mobility at room temperature due to the high optical phonon energy, which is the advantage of graphene over the other semiconductor materials for future applications.

## 5. Summary

In summary, we electrostatically realized a Dirac fermion reflector by using ballistic sawtooth-gate npn junctions. We compared our experimental results with simulations and established that the phenomenon is due to the collimation and reflection of ballistic carriers at pn interfaces. Our results constitute a vital first step toward the realization of Dirac fermion optics for functional devices such as graphene transistors with ultrahigh mobility[17–22].

## Appendix A: Dip position movement in ($n_1$, $n_2$) plane

In order to check the quantitative agreement between experiment and simulation, we plotted the dip positions of the results in Figs. 3(a) and 3(b) on the ($n_1$, $n_2$) plane [Fig. 4]. The black circles and red curves show the experimental and simulation dip positions, respectively.

## Appendix B: Results from other devices with different tilt angle

We fabricated four devices with tilt angle $\alpha \sim 45°$, but we could not observe any resistance enhancement from any of them. For example, the results and device structure of one of the devices is shown in Fig. 5. This is somehow inconsistent with Wilmart *et al*.[22]. Nevertheless, we found that some other calculation results suggest that resistance enhancement across a tilted junction, which admittedly does not have the exact same shape as our junction, shows a transmission *maximum* around the tilt angle ~ 45°, if the influence of edge scattering is considered[29]. Our next step should be to establish the best tilt angle while taking into account both our device structure and the effect of edge scattering.

**Appendix C: Temperature dependence of resistance enhancement**

Fig. 6(a) shows the temperature dependence of resistance enhancement we observed. In order to quantify the result, Fig. 6(b) shows the temperature dependence of the resistance difference between np$^-$n and np$^+$n regimes. Because of the temperature robustness of ballistic transport in graphene, the enhancement is sustained up to 280 K. The carrier mean free path $L_{\mathrm{mfp}}$ of a device with similar quality from other reports[1] is also plotted in Fig. 6(c). At 280 K and 150 K (black vertical lines), when $\varDelta R$ becomes zero and saturated to ~ 0.15 kΩ, respectively, $L_{\mathrm{mfp}}/l$ is ~ 2 and ~ 6. These numbers are reasonable as one needs at least a roundtrip inside the sawtooth-shaped gated region to achieve total reflection.

**Acknowledgments**

The authors acknowledge Rai Moriya for valuable discussions. This work was partly supported by the following: the Core Research for Evolutional Science and Technology (CREST), the Japan Science and Technology Agency (JST); JSPS KAKENHI Grant Numbers JP25107003, JP25107004, JP15K21722, JP26248061, and JP16H00982; Research Grant from the Murata Scientific Foundation; and the Project for Developing Innovation Systems of the Ministry of Education, Culture, Sports, Science, and Technology (MEXT). The research leading to these results has received partial funding from the European union Seventh Framework programme under grant nº 604391 Graphene Flagship. S. Morikawa acknowledges the JSPS Research Fellowship for Young Scientists.

**Figure captions**

Fig. 1.

(a) The refraction of ballistic carriers across the junction between different carrier density regions. (b) and (c) The schematics of the mechanism of transmission manipulation in triangular junctions, corresponding to the OFF state (b) and ON state (c). (d) $T(\phi)$ of smooth ($d = 120$ nm, solid) and abrupt ($d = 1$ nm, dotted) pn junctions. (e) $T(\phi)$ of smooth ($d = 120$ nm) pn (solid) and nn (dotted) junctions. Both (d) and (e) are derived using equation (10) in Cayssol *et al.*[31], and blue (red) curves correspond to the instance when the carrier transmits through the interface from lightly doped (heavily doped) to heavily doped (lightly doped) regions. The absolute value of lightly (heavily) doped carrier density is $0.9 \times 10^{11}$ cm$^{-2}$ ($2.0 \times 10^{12}$ cm$^{-2}$). (f) Simulation result of the resistance as a function of $n_2$ and $n_1$ across the triangular-shaped carrier-density-modulated region, which is depicted in the inset.

Fig. 2

(a) Schematic of the device geometry in the present work. Dashed lines represent the unit cell of our sawtooth-shaped junction. (b) Doping profile schematic by gates biases. (c) Experimental result of $R_{4t}$ as a function of $n_1$ and $n_2$.

Fig. 3

(a) and (b) Line cuts of Figs. 2(c) and 1(f) as functions of $n_2$ at $n_1 = 0.9, 1.2, …, 7.8 \times 10^{11}$ cm$^{-2}$ (top to bottom). (c) and (d) $R_{4t}$ for another sawtooth-shaped junction with tilt angle ~20° (c) and rectangular junction (d) as functions of $n_2$ at $n_1 = 0.9, 1.2, …, 4.2 \times 10^{11}$ cm$^{-2}$ (top to bottom). Though we cannot reach the high-$n_2$ region owing to the leakage problem, the signs of slopes around npn regions, compared to Fig. 3(a), are the same in (c) and different in (d). AFM images of the measured devices are shown in the insets of (a), (c), and (d). The black dotted lines represent the edge of graphene, and the blue scale bar indicates a length of 500 nm. We measured $R_{4t}$ by injecting current from S to D and measuring the voltage between $V_1$ and $V_2$. In the inset of (b), the device structure considered in the simulation are also shown.

**Figure captions (Appendix)**

Fig. 4

Dip positions in the ($n_1$, $n_2$) plane derived from the experimental results in Fig. 3(a) (black circles) and the simulation results in Fig. 3(b) (red lines).

Fig. 5

$R_{4t}$ as a function of $n_2$ at $n_1$ = 0.9, 1.2, …, 4.5 × $10^{11}$ cm$^{-2}$ (top to bottom) for a sawtooth-shaped junction with ~45° tilt angle. Inset: AFM image of this device. The black dotted lines represent the edge of graphene, and the blue scale bar indicates a length of 500 nm.

Fig. 6

(a) $R_{4t}$ as a function of $V_t$ at fixed $V_b$ = 1.5 V for $T$ = 6, 22, …, 294 K (top to bottom and offset vertically). (b) $\Delta R$ as a function of $T$, where $\Delta R$ is defined as the difference between $R_{4t}$ at $V_t$ = −8 V and $R_{4t}$ at $V_t$ = −4 V. (c) $L_{mfp}/l$ as a function of $T$. Vertical lines indicate the temperatures when $\Delta R$ becomes zero (280 K) and saturated to ~ 0.15 kΩ (150 K), respectively.

**References**

[1]     Wang L, Meric I, Huang P Y, Gao Q, Gao Y, Tran H, Taniguchi T, Watanabe K, Campos L M, Muller D A, Guo J, Kim P, Hone J, Shepard K L and Dean C R 2013 One-Dimensional Electrical Contact to a Two-Dimensional Material *Science* **342** 614–7

[2]     Dean C R, Young A F, Meric I, Lee C, Wang L, Sorgenfrei S, Watanabe K, Taniguchi T, Kim P, Shepard K L and Hone J 2010 Boron nitride substrates for high-quality graphene electronics. *Nat. Nanotechnol.* **5** 722–6

[3]     Zomer P J, Dash S P, Tombros N and van Wees B J 2011 A transfer technique for high mobility graphene devices on commercially available hexagonal boron nitride *Appl. Phys. Lett.* **99** 232104

[4]     Banszerus L, Schmitz M, Engels S, Goldsche M, Watanabe K, Taniguchi T, Beschoten B and Stampfer C 2016 Ballistic Transport Exceeding 28 μm in CVD Grown Graphene *Nano Lett.* **16** 1387–91

[5]     Uwanno T, Hattori Y, Taniguchi T, Watanabe K and Nagashio K 2015 Fully dry PMMA transfer of graphene on h -BN using a heating/cooling system *2D Mater.* **2** 41002

[6]     Mayorov A S, Gorbachev R V, Morozov S V., Britnell L, Jalil R, Ponomarenko L A, Blake P, Novoselov K S, Watanabe K, Taniguchi T and Geim A K 2011 Micrometer-scale ballistic transport in encapsulated graphene at room temperature. *Nano Lett.* **11** 2396–9

[7]     Taychatanapat T, Watanabe K, Taniguchi T and Jarillo-Herrero P 2013 Electrically tunable transverse magnetic focusing in graphene *Nat. Phys.* **9** 225–9

[8]     Morikawa S, Dou Z, Wang S-W, Smith C G, Watanabe K, Taniguchi T, Masubuchi S, Machida T and Connolly M R 2015 Imaging ballistic carrier trajectories in graphene using scanning gate microscopy *Appl. Phys. Lett.* **107** 243102

[9]     Bhandari S, Lee G-H, Klales A, Watanabe K, Taniguchi T, Heller E, Kim P and Westervelt R M 2016 Imaging Cyclotron Orbits of Electrons in Graphene *Nano Lett.* **16** 1690–4

[10]    Masubuchi S, Iguchi K, Yamaguchi T, Onuki M, Arai M, Watanabe K, Taniguchi T and Machida T 2012 Boundary Scattering in Ballistic Graphene *Phys. Rev. Lett.* **109** 36601

[11]    Agrawal N, Ghosh S and Sharma M 2013 Electron optics with dirac fermions: electron transport in monolayer and bilayer graphene through magnetic barrier and their superlattices *Int. J. Mod. Phys. B* **27** 1341003

[12]    Rickhaus P, Liu M-H, Makk P, Maurand R, Hess S, Zihlmann S, Weiss M, Richter K and Schönenberger C 2015 Guiding of Electrons in a Few-Mode Ballistic Graphene Channel *Nano Lett.* **15** 5819–25

[13]   Lee G-H, Park G and Lee H 2015 Observation of negative refraction of Dirac fermions in graphene *Nat. Phys.* **11** 925–9

[14]   Rickhaus P, Makk P, Richter K, Schönenberger C and Liu M-H 2015 Gate tuneable beamsplitter in ballistic graphene *Appl. Phys. Lett.* **107** 251901

[15]   Morikawa S, Masubuchi S, Moriya R, Watanabe K, Taniguchi T and Machida T 2015 Edge-channel interferometer at the graphene quantum Hall pn junction *Appl. Phys. Lett.* **106** 183101

[16]   Chen S, Han Z, Elahi M M, Habib K M M, Wang L, Wen B, Gao Y, Taniguchi T, Watanabe K, Hone J, Ghosh A W and Dean C R Electron optics with ballistic graphene junctions *arXiv:1602.08182*

[17]   Low T and Appenzeller J 2009 Electronic transport properties of a tilted graphene p−n junction *Phys. Rev. B* **80** 155406

[18]   Sohier T and Yu B 2011 Ultralow-voltage design of graphene PN junction quantum reflective switch transistor *Appl. Phys. Lett.* **98**

[19]   Gupta G, Abdul Jalil M Bin, Yu B and Liang G 2012 Performance evaluation of electro-optic effect based graphene transistors. *Nanoscale* **4** 6365–73

[20]   Sajjad R N and Ghosh A W 2013 Manipulating chiral transmission by gate geometry: Switching in graphene with transmission gaps *ACS Nano* **7** 9808–13

[21]   Jang M S, Kim H, Son Y-W, Atwater H A and Goddard W A 2013 Graphene field effect transistor without an energy gap. *Proc. Natl. Acad. Sci. U. S. A.* **110** 8786–9

[22]   Wilmart Q, Berrada S, Torrin D, Hung Nguyen V, Fève G, Berroir J-M, Dollfus P and Plaçais B 2014 A Klein-tunneling transistor with ballistic graphene *2D Mater.* **1** 11006

[23]   Han M, Özyilmaz B, Zhang Y and Kim P 2007 Energy Band-Gap Engineering of Graphene Nanoribbons *Phys. Rev. Lett.* **98** 206805

[24]   Nakaharai S, Iijima T, Ogawa S, Suzuki S, Li S L, Tsukagoshi K, Sato S and Yokoyama N 2013 Conduction tuning of graphene based on defect-induced localization *ACS Nano* **7** 5694–700

[25]   Oostinga J B, Heersche H B, Liu X, Morpurgo A F and Vandersypen L M K 2008 Gate-induced insulating state in bilayer graphene devices. *Nat. Mater.* **7** 151–7

[26]   Aparecido-Ferreira A, Miyazaki H, Li S-L, Komatsu K, Nakaharai S and Tsukagoshi K 2012 Enhanced current-rectification in bilayer graphene with an electrically tuned sloped bandgap. *Nanoscale* **4** 7842–6

[27]   Cheianov V V., Fal'ko V and Altshuler B L 2007 The focusing of electron flow and a Veselago lens in graphene p-n junctions. *Science* **315** 1252–5

[28]   Allain P E and Fuchs J N 2011 Klein tunneling in graphene: Optics with massless electrons *Eur. Phys. J. B* **83** 301–17

[29]   Sajjad R N, Sutar S, Lee J U and Ghosh A W 2012 Manifestation of chiral tunneling at a tilted graphene p-n junction *Phys. Rev. B* **86** 155412

[30]   Sutar S, Comfort E S, Liu J, Taniguchi T, Watanabe K and Lee J U 2012 Angle-dependent carrier transmission in graphene p-n junctions. *Nano Lett.* **12** 4460–4

[31]   Cayssol J, Huard B and Goldhaber-Gordon D 2009 Contact resistance and shot noise in graphene transistors *Phys. Rev. B* **79** 75428

[32]   Campos L C, Young A F, Surakitbovorn K, Watanabe K, Taniguchi T and Jarillo-Herrero P 2012 Quantum and classical confinement of resonant states in a trilayer graphene Fabry-Pérot interferometer. *Nat. Commun.* **3** 1239

[33]   Masubuchi S, Morikawa S, Onuki M, Iguchi K, Watanabe K, Taniguchi T and Machida T 2013 Fabrication and Characterization of High-Mobility Graphene p–n–p Junctions Encapsulated by Hexagonal Boron Nitride *Jpn. J. Appl. Phys.* **52** 110105

[34]   Grushina A L, Ki D-K and Morpurgo A F 2013 A ballistic pn junction in suspended graphene with split bottom gates *Appl. Phys. Lett.* **102** 223102

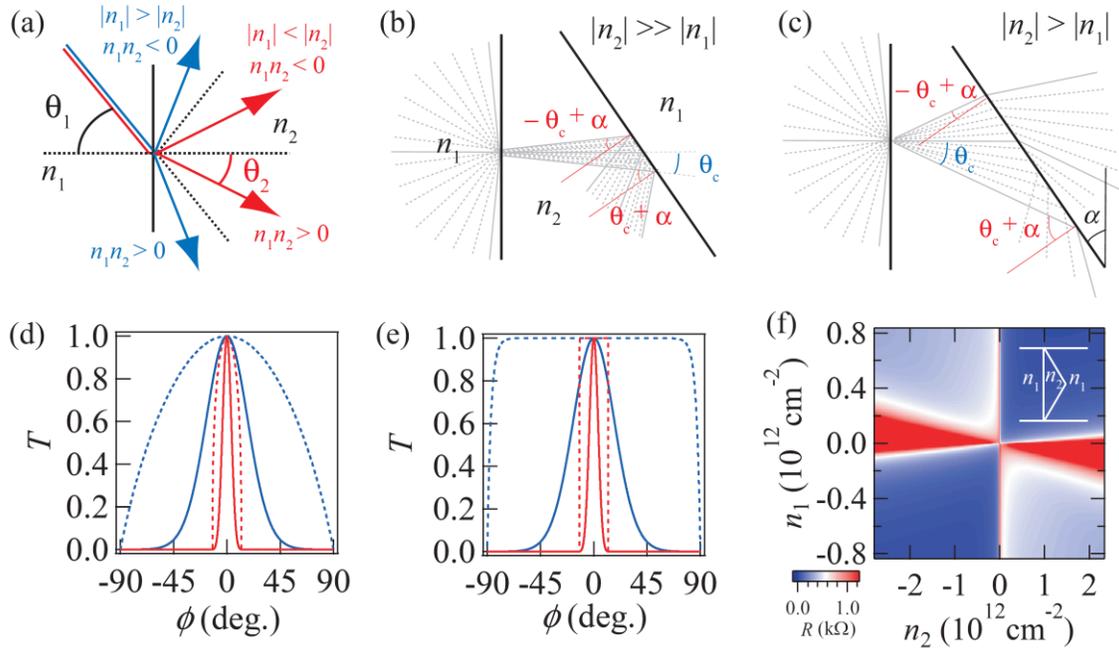

Fig. 1

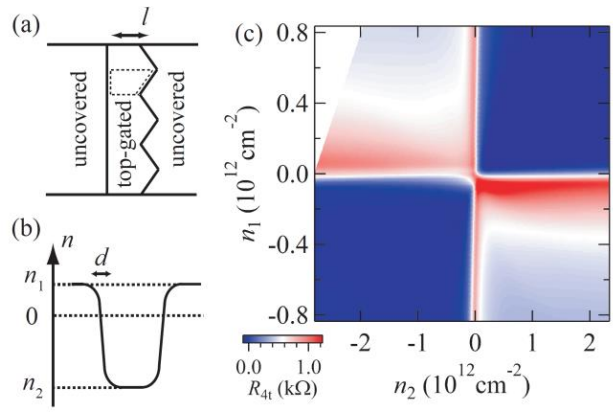

Fig. 2

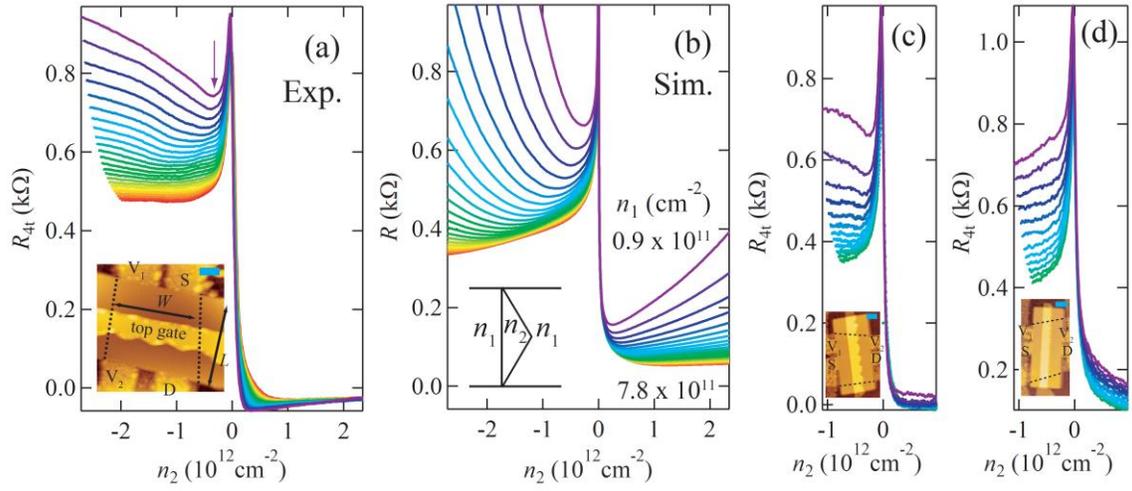

Fig. 3

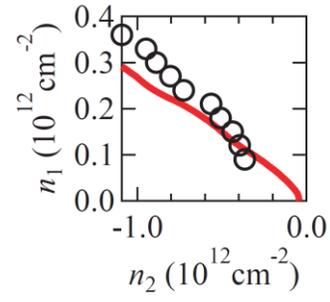

Appendix

Fig. 4

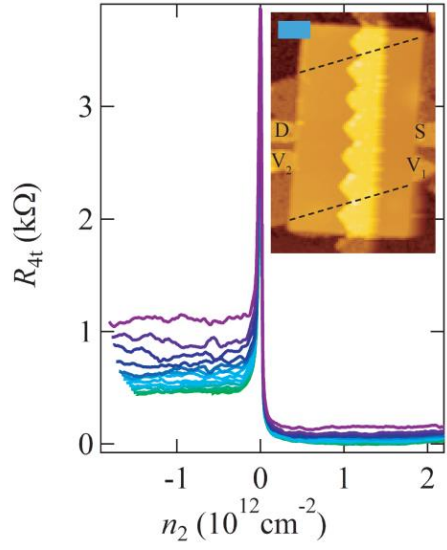

# Appendix

# Fig. 5

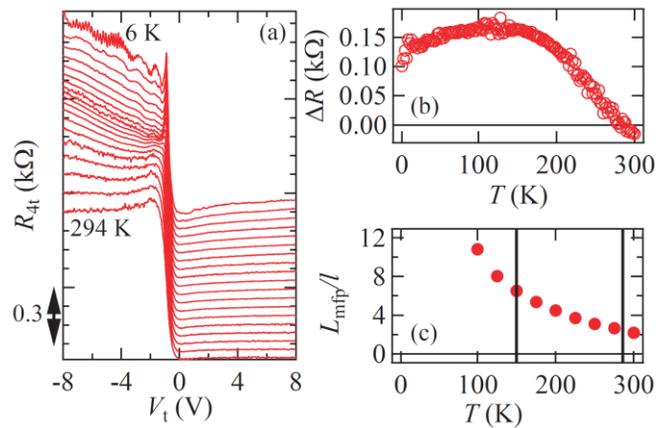

Appendix

Fig. 6